\documentstyle[epsf,12pt]{article}
\begin{document}

\catcode`@=11
\long\def\@caption#1[#2]#3{\par\addcontentsline{\csname
  ext@#1\endcsname}{#1}{\protect\numberline{\csname
  the#1\endcsname}{\ignorespaces #2}}\begingroup
    \small
    \@parboxrestore
    \@makecaption{\csname fnum@#1\endcsname}{\ignorespaces #3}\par
  \endgroup}
\catcode`@=12
\newcommand{\newc}{\newcommand}
\newc{\lat}{{\ell at}}
\newc{\one}{{\bf 1}}
\newc{\mgut}{M_{\rm GUT}}
\newc{\mzero}{m_0}
\newc{\mhalf}{M_{1/2}}
\newc{\five}{{\bf 5}}
\newc{\fivebar}{{\bf\bar 5}}
\newc{\ten}{{\bf 10}}
\newc{\tenbar}{{\bf\bar{10}}}
\newc{\sixteen}{{\bf 16}}
\newc{\sixteenbar}{{\bf\bar{16}}}
\newc{\gsim}{\lower.7ex\hbox{$\;\stackrel{\textstyle>}{\sim}\;$}}
\newc{\lsim}{\lower.7ex\hbox{$\;\stackrel{\textstyle<}{\sim}\;$}}
\newc{\gev}{\,{\rm GeV}}
\newc{\mev}{\,{\rm MeV}}
\newc{\ev}{\,{\rm eV}}
\newc{\kev}{\,{\rm keV}}
\newc{\tev}{\,{\rm TeV}}
\newc{\mz}{m_Z}
\newc{\mw}{m_W}
\newc{\mpl}{M_{Pl}}
\newc{\mh}{m_h}
\newc{\mA}{m_A}
\newc{\tr}{\mbox{Tr}}
\def\sfrac#1#2{{\textstyle\frac{#1}{#2}}}
\newc{\chifc}{\chi_{{}_{\!F\!C}}}
\newc\order{{\cal O}}
\newc\CO{\order}
\newc\CL{{\cal L}}
\newc\CY{{\cal Y}}
\newc\CH{{\cal H}}
\newc\CM{{\cal M}}
\newc\CF{{\cal F}}
\newc\CD{{\cal D}}
\newc\CN{{\cal N}}
\newc{\eps}{\epsilon}
\newc{\re}{\mbox{Re}\,}
\newc{\im}{\mbox{Im}\,}
\newc{\invpb}{\,\mbox{pb}^{-1}}
\newc{\invfb}{\,\mbox{fb}^{-1}}
\newc{\yddiag}{{\bf D}}
\newc{\yddiagd}{{\bf D^\dagger}}
\newc{\yudiag}{{\bf U}}
\newc{\yudiagd}{{\bf U^\dagger}}
\newc{\yd}{{\bf Y_D}}
\newc{\ydd}{{\bf Y_D^\dagger}}
\newc{\yu}{{\bf Y_U}}
\newc{\yud}{{\bf Y_U^\dagger}}
\newc{\ckm}{{\bf V}}
\newc{\ckmd}{{\bf V^\dagger}}
\newc{\ckmz}{{\bf V^0}}
\newc{\ckmzd}{{\bf V^{0\dagger}}}
\newc{\X}{{\bf X}}
\newc{\bbbar}{B^0-\bar B^0}
\def\bra#1{\left\langle #1 \right|}
\def\ket#1{\left| #1 \right\rangle}
\newc{\sgn}{\mbox{sgn}\,}
\newc{\m}{{\bf m}}
\newc{\msusy}{M_{\rm SUSY}}
\newc{\munif}{M_{\rm unif}}
\newc{\slepton}{{\tilde\ell}}
\newc{\Slepton}{{\tilde L}}
\newc{\sneutrino}{{\tilde\nu}}
\newc{\selectron}{{\tilde e}}
\newc{\stau}{{\tilde\tau}}
%
%
\def\NPB#1#2#3{Nucl. Phys. {\bf B#1} (19#2) #3}
\def\PLB#1#2#3{Phys. Lett. {\bf B#1} (19#2) #3}
\def\PLBold#1#2#3{Phys. Lett. {\bf#1B} (19#2) #3}
\def\PRD#1#2#3{Phys. Rev. {\bf D#1} (19#2) #3}
\def\PRL#1#2#3{Phys. Rev. Lett. {\bf#1} (19#2) #3}
\def\PRT#1#2#3{Phys. Rep. {\bf#1} (19#2) #3}
\def\ARAA#1#2#3{Ann. Rev. Astron. Astrophys. {\bf#1} (19#2) #3}
\def\ARNP#1#2#3{Ann. Rev. Nucl. Part. Sci. {\bf#1} (19#2) #3}
\def\MPL#1#2#3{Mod. Phys. Lett. {\bf #1} (19#2) #3}
\def\ZPC#1#2#3{Zeit. f\"ur Physik {\bf C#1} (19#2) #3}
\def\APJ#1#2#3{Ap. J. {\bf #1} (19#2) #3}
\def\AP#1#2#3{{Ann. Phys. } {\bf #1} (19#2) #3}
\def\RMP#1#2#3{{Rev. Mod. Phys. } {\bf #1} (19#2) #3}
\def\CMP#1#2#3{{Comm. Math. Phys. } {\bf #1} (19#2) #3}
\relax
%
%
%
\def\beq{\begin{equation}}
\def\eeq{\end{equation}}
\def\bea{\begin{eqnarray}}
\def\eea{\end{eqnarray}}
%
%
%
\newc{\ie}{{\it i.e.}}          \newc{\etal}{{\it et al.}}
\newc{\eg}{{\it e.g.}}          \newc{\etc}{{\it etc.}}
\newc{\cf}{{\it c.f.}}
\def\smuon{{\tilde\mu}}
\def\neut{{\tilde N}}
\def\char{{\tilde C}}
\def\bino{{\tilde B}}
\def\wino{{\tilde W}}
\def\higgsino{{\tilde H}}
\def\sneut{{\tilde\nu}}
%
%
%
%
\def\slash#1{\rlap{$#1$}/} 
\def\Dsl{\,\raise.15ex\hbox{/}\mkern-13.5mu D} 
\def\delsl{\raise.15ex\hbox{/}\kern-.57em\partial}
\def\Ksl{\hbox{/\kern-.6000em\rm K}}
\def\Asl{\hbox{/\kern-.6500em \rm A}}
\def\Qsl{\hbox{/\kern-.6000em\rm Q}}
\def\gradsl{\hbox{/\kern-.6500em$\nabla$}}
%
%
%
\def\bar#1{\overline{#1}}
\def\vev#1{\left\langle #1 \right\rangle}
%

\begin{titlepage}
~~
\vskip 2cm
\begin{center}
{\large\bf
Supersymmetric Flavor-Changing Sum Rules\\ as a Tool for $b\to s \gamma$
}
\vskip 1cm
{\normalsize\bf
Brian Dudley and Christopher Kolda\\
\vskip 0.5cm
{\it Department of Physics, University of Notre Dame\\
Notre Dame, IN~~46556, USA}\\[0.1truecm]
}

\end{center}
\vskip .5cm

\begin{abstract}
The search for supersymmetry (SUSY) and other classes of new physics will be tackled on two fronts, with high energy, direct detection machines, and in high precision experiments searching for indirect signatures.  While each of these methods has its own strengths, even more can be gained by finding ways to combine their results. In this paper, we examine one way of bridging these two types of experiments by calculating sum rules which link physical squark masses to the flavor-violating squark mixings.  These sum rules are calculated for minimally flavor-violating SUSY theories at both high and low $\tan\beta$. We also explore how the sum rules could help to disentangle the relative strengths of different SUSY contributions to $b\to s\gamma$, a favored channel for indirect searches of new physics. Along the way, we show that the gluino contributions to $b\to s\gamma$ can be very sizable at large $\tan\beta$.
\end{abstract}

\end{titlepage}

\setcounter{footnote}{0}
\setcounter{page}{1}
\setcounter{section}{0}
\setcounter{subsection}{0}
\setcounter{subsubsection}{0}


Over the next several years, the search for supersymmetry (SUSY) will
be advanced in two very different directions. At the LHC, searches will attempt to find evidence for direct
production of SUSY partners and to measure their masses. Meanwhile, plans are being considered for
a next generation of high-precision machines, which will look for indirect
evidence of SUSY in the $B$ system. Each program can perform
its search independent of the other, but taken together will reveal a
much richer spectrum of information about SUSY than either would
alone.

The Minimal SUSY Standard Model (MSSM) has an extremely rich structure
which can generate a multitude of phenomenologies, depending on how
SUSY is broken and how that breaking is communicated to the MSSM
sector. Even with the discovery of SUSY, it will take a
large body of data to convince ourselves that we have understood the
underlying theory. Furthermore, it seems unlikely that we can
reach this understanding without several different kinds of data.

The spectrum of the general MSSM is quite complex, involving 30
masses, 39 mixing angles, and 41 phases~\cite{Dimopoulos:1995ju}. Most of the
angles and phases are tied to the SUSY flavor sector and as such are
highly constrained already. However there exist compelling models in
which the next generation of precision measurements could uncover
SUSY flavor physics.

The masses are a different story.
It is a matter of faith among most theorists
today that the SUSY mass spectrum {\it will}\/ be found at the LHC. But to a first
approximation, the LHC is {\it only}\/ sensitive to the masses. Precision
experiments, such as LHCb, or super B-factory will be sensitive to particular
combinations of masses, angles and phases. It is vitally important
that the data from the two classes of experiments can be combined and
compared. In particular, once the LHC has measured sparticle masses,
one would like clear predictions for flavor-changing amplitudes.

In this paper, we develop a technique of sum rules for
guiding these comparisons. The basic principle for the sum rules is
simple. Within any specific model of SUSY-breaking, there are only a
small handful of independent parameters. For example, in the
much-studied minimal supergravity (mSUGRA) model, the only free
parameters are a common scalar mass ($\mzero$), a common gaugino mass
($\mhalf$), a common trilinear term ($A_0$), the bilinear term ($B_0$)
and the $\mu$-term. From these five parameters flow all 110 terms in
the MSSM Lagrangian. Obviously, then, there must exist a large number
of constraints among the terms in the Lagrangian and their
coefficients. We will derive some of these constraints as a way of
testing models, with a focus on minimally flavor-violating models.

Similar studies have been conducted before. What will set apart this
study, and make its results particularly useful, is that the result is
a set of analytic formulae which connect commonly-defined
flavor-changing parameters with experimentally-measurable masses. Only
physical masses will enter the sum rules, allowing a direct comparison
to experiment.

One of the most interesting flavor-changing neutral current (FCNC) processes is $b\to s\gamma$. This is a dimension five helicity suppressed
process in the Standard Model (SM). This allows new physics contributions to be comparable. More specifically, SUSY models can contribute
to this process through many different penguin diagrams. In the last section of this paper we will show that the
oft-neglected gluino ($\tilde{g}$) contributions can in certain regions of the parameter space become comparable to
other SUSY contributions.  In situations like this, the sum rules we derive in this paper can prove an invaluable
resource in disentangling the relative weight of the SUSY contributions.

\section{Parametrization}

Within the SM, flavor-changing currents are an
indication that the mass and interaction eigenstates for the quarks
are not exactly the same. But thanks to the very simple structure of
the quark Yukawa couplings, the only source of quark
flavor changing is the
Cabibbo-Kobayashi-Maskawa (CKM) matrix. Because the CKM matrix is
unitary, tree-level flavor-changing neutral currents (FCNCs)
are forbidden; and because it is so nearly
diagonal and most quark masses are light, loop-level FCNCs are also
highly suppressed.

But in the MSSM, there are additional sources of FCNCs at the loop
level due to the presence of the squarks. While the quarks receive
their masses only from electroweak symmetry breaking, squarks receive
masses also from SUSY breaking. The two sources need not align
and so additional rotations are necessary
in order to go from the quark to the squark mass eigenstates. For example, the $d-\tilde{d}-\tilde{g}$
interaction has the form
(in the mass eigenbasis)
$${\cal L}\sim
\left(\begin{array}{ccc} \bar d_L & \bar s_L & \bar b_L \end{array}\right)
\left(\begin{array}{ccc} ~ & ~ & ~ \\ & U^\dagger_{d_L} & \\ & & \end{array}\right)
\left(\begin{array}{ccc} ~ & ~ & ~ \\ & \tilde U_{d_L} & \\ & & \end{array}\right)
\left(\begin{array}{c} \tilde d_L \\ \tilde s_L \\ \tilde b_L \end{array}\right) \tilde g
$$
 where $U_{d_L}$ and $\tilde U_{d_L}$ are the $3\times 3$ unitary matrices which
rotate from the mass eigenbasis to the interaction eigenbasis for the
$d_L$-quarks and $\tilde{d}_L$-squarks respectively. FCNCs are generated if the
product $U_{d_L}^\dagger\tilde U_{d_L}$ is not the unit matrix.

It is actually more common in the literature to work in the
interaction basis. Here $U_{d_L}=\tilde U_{d_L}$ by definition, but
the squark mass matrices are non-diagonal. FCNCs arise
due to the presence of mass mixing insertions, $\tilde m^2
\tilde d_L^* \tilde s_L$, for example.

If one were to treat the MSSM as simply an effective theory,
one would expect the coefficients of the
flavor-changing squark mass terms to be of order
$\msusy^2$, since there is no symmetry to forbid such
mixings. (Equivalently we expect the off-diagonal terms in
$U_{d_L}^\dagger \tilde U_{d_L}$ to be $O(1)$.)
But the experimental absence of large FCNCs
indicates that these coefficients are instead very small
compared to $\msusy^2$. This problem has driven much of the
model-building activity in SUSY for the last two decades, and measuring
flavor-changing SUSY effects will play a vital role in unraveling the physics
that generates SUSY breaking.

Most realistic SUSY models prevent large FCNCs by requiring degeneracies among squarks with
identical gauge quantum numbers. Thus all $d_L$-type squarks would have the same mass, all $u_R$-type quarks
would have their same mass, and so on. If such degeneracies were perfect, there would be no new FCNCs at all.
But in every conceivable model, the degeneracies are broken. At the very least,
such relationships are not perserved by quantum corrections coming
from the Yukawa sector. However, there could be other non-degenerate contributions to the squark masses as
well (K\"ahler corrections, flavor-dependent D-terms, {\it etc.}).

Flavor-changing contributions can affect the squark mass spectrum in
two ways, by {\sl (i)} splitting the masses of the squarks, and by
{\sl (ii)} mixing the squarks. In the interaction basis, these appear respectively
as non-degeneracies among the diagonal terms in the squark
mass matrices, and off-diagonal terms in the squark mass
matrices. The first effect does not generate any new FCNCs, while the
second does, a difference which is often misunderstood.

Let us define our notation.
Before electroweak symmetry breaking, the only kinds of squark mixing
allowed would be among those with the same $SU(2)\times U(1)$ quantum
numbers. Thus the $\{\tilde{d}_L, \tilde{s}_L, \tilde{b}_L\}$ could mix, as could the
$\{\tilde{u}_R, \tilde{c}_R, \tilde{t}_R\}$ and the $\{\tilde{d}_R, \tilde{s}_R, \tilde{b}_R\}$. The $\{\tilde{u}_L,
 \tilde{c}_L, \tilde{t}_L\}$ can intermix as well, though this mixing is aligned to that in
the $\tilde{d}_L$ sector by SU(2). The mixings in these sectors we call LL or
RR mixing, since both states are either left- or right-handed. As a
consequence there will be two $3\times 3$ mass matrices in the down
sector, and two in the up sector. For example,
\beq
{\cal M}^2_{D,LL} = \left(\begin{array}{ccc}
(\tilde m^2_{D1})_{LL} & (\Delta^d_{12})_{LL} & (\Delta^d_{13})_{LL} \\
(\Delta^d_{21})_{LL} & (\tilde m^2_{D2})_{LL} & (\Delta^d_{23})_{LL} \\
(\Delta^d_{31})_{LL} & (\Delta^d_{32})_{LL} & (\tilde m^2_{D3})_{LL}
\end{array}\right)
\eeq
with similar matrices for the $\tilde{d}_R$, $\tilde{u}_L$ and $\tilde{u}_R$ sectors.

After electroweak symmetry
breaking, left-right mixing is allowed, and
the LL and RR squark mass matrices combine to form two
$6\times6$ matrices, one each for $\tilde{u}$- and $\tilde{d}$-squarks,  in which all possible LL, RR and LR mixings are allowed.
We can parametrize the mixing by the matrix:
\beq
\tilde M^2_{D} = \left(\begin{array}{cc}
{\cal M}^2_{D,LL} & {\cal M}^2_{D,LR} \\
{\cal M}^2_{D,RL} & {\cal M}^2_{D,RR}
\end{array}\right).
\eeq
Because the matrix is hermitian, so are
${\cal M}^2_{D,LL}$ and ${\cal M}^2_{D,RR}$, while ${\cal M}^2_{D,LR}
= \left({\cal M}^2_{D,RL}\right)^\dagger$.

In order to make this parametrization useful, we must specify a basis
in which the masses and their mixings are to be calculated.
In writing the superpotential of the MSSM, it is possible to rotate
the $\hat Q$, $\hat U$ and $\hat D$ superfields in order to diagonalize the
$d$-sector Yukawa couplings, but not simultaneously the
$u$-sector. This means that the general Yukawa interactions can be
written as:
\beq
W=\hat Q_i (V^\dagger Y_U)_{ij} \hat U_j \hat H_u + \hat Q_i Y_{D,ij} \hat D_j \hat H_d
\eeq
where $Y_U$ and $Y_D$ are {\it diagonal}\/ $3\times 3$ Yukawa matrices and $V$ is the
usual CKM matrix. The $\hat Q_i$ fields are not the mass
eigenstates, but can be identified with the eigenstates of the weak
interactions. We can define the {\it quark}\/ mass eigenstate by a rotation on
its left-handed component: $u_{L,i} = V^*_{ij} Q_{u,j}$ where
$Q$ is the fermionic part of the superfield $\hat Q$. Then in order
to maintain flavor-diagonal gluino interactions it is also necessary
to rotate the $\tilde u_L$ quarks by the same amount:
$\tilde u_{L,i} = V^*_{ij} \tilde Q_{u,j}$. It is this basis that is commonly called the
``superCKM basis" and it is
in this basis that there is a clear and simple connection between the mass insertions and
the amplitudes for various flavor-changing processes.

In the superCKM basis, each $3\times 3$ submatrix of the squark
$6\times6$ mass matrix can be expressed in a simple form:
\bea
{\cal M}^2_{D,LL} & = & \tilde m^2_{Q} + D_{d_L} +
m_D^2 \nonumber\\
{\cal M}^2_{D,RR} & = & \tilde m^2_{D} + D_{d_R} + m_D^2 \nonumber\\
{\cal M}^2_{D,LR} & = & A_D\, v_d + \mu m_D\tan\beta \nonumber\\
{\cal M}^2_{U,LL} & = & V\tilde m^2_{Q}V^\dagger + D_{u_L} +
m_U^2 \label{Matmass}\\
{\cal M}^2_{U,RR} & = & \tilde m^2_{U} + D_{u_R} + m_U^2 \nonumber\\
{\cal M}^2_{U,LR} & = & V A_U\, v_u + \mu m_U\cot\beta . \nonumber
\eea
In the above expressions, the $D_{q_{L,R}}$ are the flavor-diagonal
$D$-term contributions, and $m_U$ and $m_D$ are the flavor-diagonal
quark mass matrices. The $A_U$ and $A_D$ are $3\times3$ trilinear mass
matrices.
The terms $\tilde m^2_{Q}$, $\tilde m^2_{D}$ and $\tilde m^2_{U}$ are soft mass
terms which have been run from the SUSY-breaking scale to the weak
scale.

The only difficulty in thinking about the superCKM basis is that the
rotations necessary to get into this basis are not
SU(2)-invariant. In order to make our notation clear, we will define
some rotated mass terms:
\beq
\tilde m_{D_L}^2 = \tilde{m}^2_Q,
\quad\quad \tilde m_{U_L}^2 = V \tilde m^2_Q V^\dagger, \quad\quad  \bar A_U = V A_U.
\eeq
Because we will treat $\tilde m^2_{U_L}$ and $\tilde m^2_{D_L}$ as
separate parameters, it will appear that we are explicitly breaking
SU(2) invariance; in fact SU(2) invariance above the weak scale
will always be preserved, even if it is hidden in the equations.

Of the terms that appear in Equations (\ref{Matmass}) only a subset can
generate FCNCs. In particular, the $D$-terms, and the quark mass
and $\mu$-contributions (which both come from $F$-terms) are
always flavor diagonal. The non-diagonal entries come from the soft scalar
mass terms and the $A$-terms. Since we are already working in the
superCKM basis, it is these off-diagonal terms which can be
immediately used in calculations of FCNC processes.

\section{Flavor-Changing Sum Rules in Degeneracy\\ Models}

Among the classic techniques for approaching the SUSY flavor problem (namely degeneracy, decoupling and
alignment), it is degeneracy that dominates most model-building efforts.
When one speaks of a model as exhibiting degeneracy, several different ideas might be meant.
In the most extreme cases, degeneracy refers to the complete (and therefore unrealistic) degeneracy of all
squarks in the MSSM. Such a degeneracy is immediately broken by gauge and Yukawa interactions, and as most easily
observed in the renormalization group running of the sparticle masses. A slightly more realistic version is complete
degeneracy of all squarks (and sleptons) at some ultraviolet scale, usually taken to be the gauge unification
scale or string scale. Typical among such models would be the canonical minimal supergravity (mSUGRA) model,
which is very often studied in its realization as the constrained minimal supersymmetric standard model (CMSSM).

But degeneracy models need not possess degeneracy between the sleptons and squarks, or even among the
various squarks, so long as squarks with identical gauge quantum numbers are degenerate. Thus in the most
general degeneracy model, there are 2 independent masses in the slepton sector and 3 in the squark sector,
all without introducing any new source of flavor changing. These separate degeneracies are assumed to hold at
some scale in the ultraviolet, and then radiative corrections due to Yukawa interactions split the degeneracies in the
infrared. Such models are therefore ``minimally flavor violating" (MFV); all the flavor violation comes from the Yukawa
couplings, mimicking the structure of the Standard Model~\cite{D'Ambrosio:2002ex,Buras:2003jf}. This broader definition of degeneracy
includes not only mSUGRA but also gauge-mediated, anomaly-mediated, gluino-mediated and most other commonly
studied models of the MSSM.

Though the degeneracy models form a preferred class of models, previous discussions of how to relate
flavor-changing rates to LHC observables usually take place within the confines of only one or another
particular model. For example, FCNCs in the context of mSUGRA are very well studied
~\cite{Gabbiani:1996hi,Gabbiani:1988rb,Bertolini:1990if}.
If and when SUSY particles are discovered at the Tevatron or LHC, a great many theorists will take whatever physical masses
have been  measured, translate them into running ($\overline{\mbox{MS}}$ or $\overline{\mbox{DR}}$) masses,
run them up to the GUT scale using the RGEs of the MSSM, define a range of unified
parameters ($M_0$, $A_0$, $M_{1/2}$) consistent with the data, then run these back down to determine
allowed ranges for a host of other observables. These will be used to motivate, or compare to, findings at LHCb or
other high-precision flavor experiments.

This default procedure has several problems. First, it must be done model by model, so that anomaly-mediated
models must be treated separately from mSUGRA models. Second, it compounds the experimental uncertainties on
the measured sparticle masses. The renormalization group running, the matching onto ultraviolet boundary
conditions, and the running back down all bring in new sources of error which magnify the
error bars on the original data. Third, the usual procedure requires as input non-physical parameters,
such as $\tan\beta$ and $A_t$, which may not be available, and without which it is nearly impossible to
predict FCNC amplitudes within mSUGRA.

We propose to shortcut this lengthy process and go almost directly from
measured masses to flavor-changing amplitudes by using sum
rules. The sum rules will directly give the off-diagonal flavor-changing mass insertions in terms of the measured mass eigenvalues.
These sum rules encode the boundary conditions from the high
scale {\it and}\/ the RGE flow of the soft parameters in such a way as
to make them invisible during the calculation. This method will have
many advantages, and a few disadvantages, over the traditional method
described above.

The main advantage is it creates a path directly connecting the
measured masses from the LHC to constraints on FCNCs measured at
precision machines.  Errors and ambiguities in the running are
already taken into account and cancelled. The precise UV boundary conditions are irrelevant because the sum
rules will test the idea of degeneracy itself, not a specific version of it. And because the sum rules will only
involve physically measured masses, uncertainties in $\tan\beta$ or $A_t$ are minimized.

It also will test degeneracy models as a class, rather than individually. This is of course also one of its
disadvantages: different degeneracy models predict different amounts of flavor changing, and these sum rules
will not provide a means for differentiating among the degeneracy models.

Another key disadvantage will be the number of masses which will be need to be measured in order to
use the sum rules.  The number is not particularly high if $\tan\beta$ is low, but gets more cumbersome
as $\tan\beta$ becomes large, as we will see.  Unfortunately it will be these high $\tan\beta$ sum rules that will play a key role in unlocking the
size of the gluino contribution $b\to s \gamma$, as it is at high $\tan\beta$ for which the gluino contributions
become important.

What the sum rules cannot eliminate is the need to translate physical (on-shell or pole) masses into running masses. The sum rules given here will be given in the $\overline{\mbox{MS}}$ (or, since we only work at one loop, $\overline{\mbox{DR}}$) scheme. This is an unavoidable issue, but one which is slightly ameliorated by the form of the sum rules themselves, as we will discuss.

In spirit, this idea is similar to that of Martin and Ramond~\cite{Martin:1993ft}
who used the boundary conditions of mSUGRA along with relations among
the RGEs to derive sum rules among the squark and slepton
masses. Because the Martin and Ramond sum rules assumed mSUGRA
boundary conditions, they are in fact a nice check on the mSUGRA
ansatz. In the end, our sum rules will also provide a check on the mass degeneracy
ansatz, and also on the implicit assumption that no new source of flavor
physics is present at scales below the unification scale.

\subsection{Renormalization Group Evolution}

In the models we will be considering, all scalar masses, with the same quantum numbers, are degenerate at some high scale.
In running the scalar masses down from the high scale to the
weak scale, the flavor-changing already present in the Yukawa matrices
is imprinted in the scalar mass spectrum. Because the underlying
source of flavor changing is the Yukawa couplings (and therefore the
CKM matrix), these models are minimally flavor violating~\cite{D'Ambrosio:2002ex,Buras:2003jf}.

The RGEs for the soft mass matrices are well known~\cite{Martin:1993zk}
and we will work with them only to first order:
\bea
16\pi^2\frac{d\tilde m_{Q}^2}{dt} &=&
(\tilde m^2_Q + 2m^2_{H_u}\one) V^\dagger Y_UY_U^\dagger V
+(\tilde m^2_Q+2m^2_{H_d}\one) Y_DY_D^\dagger
\nonumber\\
&& +\left[V^\dagger Y_UY_U^\dagger V+Y_DY_D^\dagger\right] \tilde m^2_Q
+2 V^\dagger Y_U\tilde m_U^2Y_U^\dagger V +2Y_D \tilde m_D^2 Y_D^\dagger
\nonumber\\
&& +2 A_UA_U^\dagger + 2A_D A_D^\dagger -\left(\sfrac{32}{3} g_3^2 M_3^2 +
6g_2^2M_2^2+\sfrac{2}{15}g_1^2M_1^2 \right)\one,
\nonumber\\
16\pi^2\frac{d\tilde m^2_U}{dt} &=&
(2\tilde m^2_U+4m^2_{H_u}\one)Y_U^\dagger Y_U + 4 Y_U^\dagger \tilde
m^2_Q Y_U + 2Y_U^\dagger Y_U\tilde m_U^2 + 4A_U^\dagger A_U \nonumber\\
&& -\left(\sfrac{32}{3}g_3^2M_3^2 +\sfrac{32}{15} g_1^2M_1^2\right)\one,
\nonumber\\
16\pi^2\frac{d\tilde m^2_D}{dt} &=&
(2\tilde m^2_D+4m^2_{H_d}\one)Y_D^\dagger Y_D + 4 Y_D^\dagger \tilde
m^2_Q Y_D + 2Y_D^\dagger Y_D\tilde m_D^2 + 4A_D^\dagger A_D \nonumber\\
&& -\left(\sfrac{32}{3}g_3^2M_3^2 +\sfrac{8}{15} g_1^2M_1^2\right)\one,
\\
16\pi^2\frac{dA_U}{dt} &=&
\left[6\tr(Y_U^\dagger V A_U) + 4A_UY_U^\dagger V +2 A_DY_D^\dagger
+\sfrac{32}{3} g_3^2 M_3 + 6g_2^2 M_2 + \sfrac{26}{15}g_1^2 M_1\right]
V^\dagger Y_U \nonumber \\
& &+\left[3\tr(Y_U^\dagger Y_U) + 5V^\dagger Y_U Y_U^\dagger V+ Y_DY_D^\dagger
-\sfrac{16}{3} g_3^2 -  3g_2^2 -\sfrac{13}{15}g_1^2\right]A_U,
\nonumber\\
16\pi^2\frac{dA_D}{dt} &=&
\left[6\tr(Y_D^\dagger A_D) + 4A_DY_D^\dagger +2 A_UY_U^\dagger V
+\sfrac{32}{3} g_3^2 M_3 + 6g_2^2 M_2 + \sfrac{14}{15}g_1^2
M_1\right]Y_D \nonumber \\
& &+\left[3\tr(Y_D^\dagger Y_D) + 5Y_D Y_D^\dagger + V^\dagger Y_UY_U^\dagger V
-\sfrac{16}{3} g_3^2 -  3g_2^2 -\sfrac{7}{15}g_1^2\right]A_D,\nonumber
\eea
where $t=\log Q^2$. Because we have defined $Y_U$ and $Y_D$ to be
diagonal, factors of $V$ appear scattered throughout. One should
interpret these as the ``running'' CKM matrix, $V(Q)$, or perhaps
more simply, as the matrix which diagonalizes the $Y_U$ matrix at a
given scale $Q$.

Once we go to the superCKM basis,
the RGE for $\tilde m^2_Q$ needs to
be reexpressed. In this basis, the RGE for $\tilde m^2_{D}$ remain unchanged while others are
rotated to a degree:
\bea
16\pi^2\frac{d\tilde m_{D_L}^2}{dt} &=&
(\tilde m^2_{D_L} + 2m^2_{H_u}\one) V^\dagger Y_U^2 V
+(\tilde m^2_{D_L}+2m^2_{H_d}\one) Y_D^2
\nonumber \\
& &+\left[V^\dagger Y_U^2 V+Y_D^2\right] \tilde
m^2_{D_L}
+2 V^\dagger Y_U\tilde m_U^2Y_U^\dagger V +2Y_D \tilde m_D^2 Y_D^\dagger
\nonumber \\
& &+2 V^\dagger \bar A_U\bar A_U^\dagger V + 2A_D A_D^\dagger +
\mbox{gauge terms},\nonumber\\
16\pi^2\frac{d\tilde m_{U_L}^2}{dt}&=&
(\tilde m^2_{U_L} + 2m^2_{H_u}\one) Y_U^2
+(\tilde m^2_{U_L}+2m^2_{H_d}\one)V Y_D^2 V^\dagger
\nonumber\\
& &+\left[Y_U^2 +VY_D^2 V^\dagger\right] \tilde m^2_{U_L}
+2 Y_U\tilde m_U^2Y_U^\dagger +2VY_D \tilde m_D^2 Y_D^\dagger V^\dagger
\nonumber \\
& &+2 \bar A_U\bar A_U^\dagger + 2VA_D A_D^\dagger V^\dagger
+ \mbox{gauge terms},
\\
16\pi^2\frac{d\bar A_U}{dt} &=&
\left[6\tr(Y_U^\dagger \bar A_U) + 4\bar A_UY_U^\dagger  +2 VA_DY_D^\dagger
V^\dagger +\mbox{gauge terms}
\right] Y_U \nonumber \\
& & +\left[3\tr(Y_U^\dagger Y_U) + 5Y_U Y_U^\dagger + VY_DY_D^\dagger V^\dagger
+\mbox{gauge terms} \right]\bar A_U,\nonumber\\
16\pi^2\frac{dA_D}{dt} &=&
\left[6\tr(Y_D^\dagger A_D) + 4A_DY_D^\dagger
+2 V^\dagger \bar A_UY_U^\dagger V
+\mbox{gauge terms}
\right]Y_D \nonumber\\
& & +\left[3\tr(Y_D^\dagger Y_D) + 5Y_D Y_D^\dagger + V^\dagger Y_UY_U^\dagger V
+\mbox{gauge terms}
\right]A_D.\nonumber
\eea
Here we are rewriting $Y_QY_Q^\dagger$ as $Y_Q^2$ since all Yukawa
matrices are defined to be diagonal and real. Note also that the RGEs for $\tilde{m}_{D_L}^2$ and
$\tilde{m}_{U_L}^2$ are not independent; one is simply the other rotated into the superCKM basis.

What happens when we apply degeneracy boundary conditions at the high scale? The usual
degeneracy conditions that $\tilde m_Q^2,\tilde m_U^2,\tilde m_D^2
\propto\one$ implies that the two left handed squark masses unify to the same value at the
 the boundary scale, $\tilde m_{U_L}^2 =\tilde m_{D_L}^2
 =\tilde{m}_Q^2$. We are also enforcing another boundary condition that the $A$-terms
be proportional to the superpotential couplings, so that
$A_D=A_{D_0} Y_D$ and $A_U=A_{U_0} V^\dagger Y_U$ with $A_{U_0},A_{D_0}$  dimensionful
constants; the $A_U$ relation implies
$\bar A_U=A_{U_0} Y_U$, which is diagonal. This assures that the the flavor mixing continues to come only from the CKM
matrix, consistent with MFV.

Rather than solve the RGE's in one step, consider their approximate
solution at a scale $t_0+\delta t$ close to the unification scale, $t_0$.
Then the change in the soft mass terms can be written as:
\bea
16\pi^2\frac{\delta\tilde m_{U_L}^2}{\delta t} &=&
2\left(\tilde m^2_{Q0} + \tilde m^2_{U0} + \tilde m^2_{H_u0}
+ A_{U0}^2\right)Y_U^2
\nonumber\\
 & &+2\left(\tilde m^2_{Q0} + \tilde m^2_{D0} + \tilde m^2_{H_d0}
+ A_{D0}^2\right)VY_D^2 V^\dagger+\mbox{(gauge terms)},
\nonumber\\
16\pi^2\frac{\delta\tilde m_{D_L}^2}{\delta t} &=&
2\left(\tilde m^2_{Q0} + \tilde m^2_{U0} + \tilde m^2_{H_u0}
+ A_{U0}^2\right)V^\dagger Y_U^2 V
\nonumber\\
 & &+2\left(\tilde m^2_{Q0} + \tilde m^2_{D0} + \tilde m^2_{H_d0}
+ A_{D0}^2\right)Y_D^2+\mbox{(gauge terms)},
\nonumber\\
16\pi^2\frac{\delta\tilde m^2_U}{\delta t} &=&
4\left(3m_0^2+A_0^2\right)Y_U^2
+ \mbox{(gauge terms)},
\\
16\pi^2\frac{\delta\tilde m^2_D}{\delta t} &=&
4\left(3m_0^2+A_0^2\right) Y_D^2 +\mbox{(gauge terms)},
\nonumber\\
16\pi^2\frac{\delta\bar A_U}{\delta t} &=&
\left[9A_0\tr(Y_U^2)+9A_0Y_U^2+3A_0VY_D^2V^\dagger
+\mbox{gauge terms}\right]Y_U,
\nonumber\\
16\pi^2\frac{dA_D}{dt} &=&
\left[9A_0\tr(Y_D^2)+9A_0Y_D^2+3A_0V^\dagger Y_U^2 V
+\mbox{gauge terms}\right]Y_D. \nonumber
\eea
Because we choose $\delta t$ small, we can take $Y_U,
Y_D\mbox{ and }V$ to be their high scale values.

It is clear that in the superCKM basis, off-diagonal mass terms are
generated for the left-handed squarks, but not for the
right-handed. The off-diagonal terms in $\delta\tilde m^2_{U_L}$ are
proportional to $VY_D^2V^\dagger$ and are
therefore only important at large $\tan\beta$ when the bottom Yukawa
coupling becomes large. However the off-diagonal terms in
$\delta\tilde m^2_{D_L}$ are proportional to $V^\dagger Y_U^2 V$
which brings in the large top quark Yukawa. Thus one finds
 the well-known result that in the mSUGRA models, the
leading flavor-changing mass insertions are left-handed and
are proportional to $V^\dagger
Y_U^2 V$, echoing the structure of the flavor-changing operators in
the Standard Model itself.

A short discussion of the $A$-terms is in order and will be useful
later. It is customary to think of the $A$-terms in an mSUGRA model as
having the form $A=a Y$ where $a$ is a dimensionful parameter that runs
according to its own RGE, and $Y$ is the appropriate Yukawa matrix.
While the mSUGRA boundary conditions do
yield this form (\ie, $A_U(M_X) = A_0 Y_U(M_X)$), this form is {\it
  not}\/ preserved by the RGEs. At any scale $Q$ below the
unification scale we can only speak about the dimensionful $3\times 3$
matrices $A_{U,D}(Q)$ which are no longer strictly proportional to
$Y_{U,D}(Q)$. At large $\tan\beta$ this will generate some
difficulties in finding a sum rule as we will discuss in Section 2.3.

We will now examine the RGEs for the mass matrices and from them extract the
off-diagonal, flavor-mixing elements in the superCKM basis, along with the squark mass eigenvalues.
We will then solve, with a bit of algebra, for the relationship
between the the off-diagonal elements and the eigenvalues, presenting the result in terms of sum rules.  These sum rules will be the link between the FCNC-generating $\Delta$'s and the physical masses as measured at the LHC.

\subsection{Sum Rules: The Low $\tan\beta$ Case}

At a scale only $\delta t$ away from the unification scale, the soft
mass-squared parameters all take the values $\tilde m^2_i = m_{i,0}^2 +
\delta \tilde m^2_i$, where the $\delta\tilde m^2_i$ were defined in
the previous section.
In order
to find the actual physical
mass eigenvalues we need to diagonalize the left-handed sector (the
right-handed being already diagonalized). The non-universal
corrections to $\tilde m^2_{U_L}$ have the form
\(Y^{2}_{U}+VY^{2}_{D}V^{\dagger} \).
Since we are working right now at low $\tan\beta$, we drop all $Y_D$
terms and all terms in $Y_U$ other than $y_t$. The matrix
controlling the flavor changing in the $\tilde{u}_L$ sector is then
$Y_U^2 = \mbox{diag}(0,0,y_t^2)$. Thus the left-handed stop will receive large corrections to its
mass which split it from the other squarks, but this will not by itself
lead to any gluino-mediated FCNCs.

In the $\tilde{d}_L$ sector things are somewhat different. Here, the relevant
non-universal terms have the form
\(V^{\dagger}Y^{2}_{U}V+Y^{2}_{D} \). Setting all Yukawas other than
$y_t$ to zero yields an hermetian matrix of the form:
\beq
V^\dagger Y_U^2 V = y_t^2 \left(\begin{array}{ccc}
|V_{31}|^2 &  V_{31}^*V_{32} & V_{31}^* V_{33} \\
 & |V_{32}|^2 & V_{32}^* V_{33} \\
 & & |V_{33}|^2 \end{array}
\right), \label{VCKM}
\eeq
where $V_{31} = V_{td}, V_{32} = V_{ts}$ and so on.
The eigenvalues of this matrix are obviously $\{0,0,y_t^2\}$.

Working in the superCKM basis, we can write down the squark mass matrices at the scale
$t+\delta t$:
\bea
{\cal M}^2_{U,LL} &=& (\alpha_Q+D_{u_L})\one + (\gamma_u+v_u^2) Y_U^2,\nonumber \\
{\cal M}^2_{D,LL} &=& (\alpha_Q+D_{d_L})\one + \gamma_u\, V^\dagger Y_U^2V,\nonumber \\
{\cal M}^2_{U,RR} &=& (\alpha_U+D_{u_R})\one + (2\gamma_u+v_u^2) Y_U^2,\label{Masses} \\
{\cal M}^2_{D,RR} &=& (\alpha_D+D_{d_R})\one,\nonumber
\eea
where
\bea
\alpha_Q &=& \tilde{m}_Q^2 - (\delta t/16\pi^2) \left(\frac{32}{3} g_3^2 M_3^2
+6g_2^2M_2^2+\frac{2}{15}g_1^2M_1^2\right),\nonumber \\
\alpha_U &=& \tilde{m}_U^2 - (\delta t/16\pi^2)\left(\frac{32}{3} g_3^2 M_3^2
+\frac{32}{15}g_1^2M_1^2\right),\nonumber \\
\alpha_D &=& \tilde{m}_D^2 - (\delta t/16\pi^2)\left(\frac{32}{3} g_3^2 M_3^2
+\frac{8}{15}g_1^2M_1^2\right), \\
\gamma_u &=& -(\delta t/8\pi^2)\left(\tilde m_{Q}^2 +
\tilde m_{U}^2 + m_{H_u}^2 + 2A_U^2\right),\nonumber \\
\gamma_d &=& -(\delta t/8\pi^2)\left(\tilde m_{Q}^2 +
\tilde m_{D}^2 +  m_{H_d}^2 + 2A_D^2\right).\nonumber
\eea
In the above expressions all masses are evaluated at the scale $t_0$.

We begin by setting the LR mixing terms to zero in order to isolate generational mixing; the LR mixing
will be put back in later in the calculation.  The mass matrices are then easily diagonalized and yield the
following mass eigenvalues:
\bea
(\tilde m_{U_{L}}^{2})_{1,2} & = & \alpha_Q + D_{u_L},
\nonumber \\
(\tilde m_{U_{L}}^{2})_{3} & = & \alpha_Q +D_{u_L} +(\gamma_u+v_u^2) y_t^2,
\nonumber \\
(\tilde m_{U_{R}}^{2})_{1,2} & = & \alpha_U + D_{u_R},
\nonumber \\
(\tilde m_{U_{R}}^{2})_{3} & = &  \alpha_U + D_{u_R} + (2\gamma_u+v_u^2) y_t^2,
\\
(\tilde m_{D_{L}}^{2})_{1,2} & = & \alpha_Q + D_{d_L},
\nonumber \\
(\tilde m_{D_{L}}^{2})_{3} & = & \alpha_Q + D_{d_L} + \gamma_u y_t^2,
\nonumber \\
(\tilde m_{D_R}^2)_{1,2,3} & = & \alpha_D + D_{d_R}.\nonumber
\eea
These masses require some interpretation. Though for now they are evaluated at some scale $t$ close to $t_0$, we will eventually continue the running down to the SUSY mass scale. At that scale, these masses are almost the physically observable masses, the difference between the physical masses and these being LR mixing and threshold loop corrections.

The next step is to evaluate the off-diagonal elements which are
related to the various mixings. Looking at Eqs.~(\ref{Masses}) it
can be seen that there is no off-diagonal flavor mixing in the
RR sector, nor in the LL up sector. That is, at low $\tan\beta$
mass degeneracy models give:
\beq
(\Delta^u_{ij})_{LL} = (\Delta^u_{ij})_{RR} = (\Delta^d_{ij})_{RR} = 0,
\eeq
which is well known.
For the LL down sector, mixing is induced by the top Yukawa
coupling,
\beq
(\Delta^d_{ij})_{LL} = \gamma_u\,V^*_{3i}V_{3j}y_t^2.
\eeq
The expression for $(\Delta^d_{ij})_{LL}$ is given in terms of two measurable parameters about which we know much (the CKM elements), one parameter which can be extracted from data once $\tan\beta$ is known ($y_t$) and another which is completely unphysical ($\gamma_u$). Yet those same parameters also occur in the physical masses and can be extracted from them. Unfortunately, the physical masses also include a number of new parameters ($\alpha_i$, $D_i$) that do not appear in the $\Delta$-term. Therefore, if we wish to extract $\gamma_u$ and $y_t$ from the physical masses, we will need to find combinations of physical masses which do not depend on these new unphysical parameters.

Luckily such combinations are easy to find.
After a little bit of algebra the following relation (``sum rule") is found:
\bea
(\Delta^d_{ij})_{LL} &=& V^*_{3i}V_{3j}\left[
(\tilde m^2_{D_L})_3+ (\tilde m^2_{D_R})_3 -
(\tilde m^2_{D_L})_1- (\tilde m^2_{D_R})_1\right] \nonumber \\
&=& V^*_{3i}V_{3j}\left[
\tilde m^2_{b_1}+ \tilde m^2_{b_2} -
\tilde m^2_{d_L} - \tilde m^2_{d_R}\right],
\label{sum1}
\eea
where in the second line we have re-expressed the mass eigenstates in the more standard notation.

A couple comments are now necessary. In the above sum rule, the mass eigenstates for the first two generation of squarks are designated by their chirality; in actuality, these are not pure left- or right-handed states, but because LR mixing in the first two generations is minimal in these models, we can presume to label by chirality anyway. But this is not true in the third generation, where we use the labels
$\tilde{b}_i$ (i = 1,2) for the two sbottom mass eigenstates.

One should also note that we have derived the sum rule at a scale $t$ close to $t_0$, far from the physical mass scale. Nonetheless, the sum rule itself is scale invariant. Both sides renormalize identically and so we can evaluate the formula at any scale, including the SUSY scale where the masses correspond to measurable observables.

Of course, a number of sum rules can be written which express the same
physics. For example, the above sum rule can just as well be written
as:
\bea
(\Delta^d_{ij})_{LL} &=& \frac13 V^*_{3i}V_{3j}\left[
(\tilde m^2_{U_L})_3+ (\tilde m^2_{U_R})_3 - 2m_t^2
-(\tilde m^2_{U_L})_1+ (\tilde m^2_{U_R})_1\right] \nonumber \\
&=& \frac13 V^*_{3i}V_{3j}\left[
\tilde m^2_{t_1}+ \tilde m^2_{t_2} - 2m_t^2 -
\tilde m^2_{u_L} + \tilde m^2_{u_R}\right],
\label{sum2}
\eea
if we happen to have data on the stops rather than the sbottoms. The specific sum rule one chooses to use depends on the data at hand, though using several sum rules does provide a consistency check on the mass unification assumption.

What happens when we put the LR mixing back into the calculation? The
general forms of the LR mixing terms in the superCKM basis are:
\bea
{\cal M}^2_{U,LR} &=& \bar A_U v_u + Y_U\mu v_d, \\
{\cal M}^2_{D,LR} &=& A_D v_d + Y_D\mu v_u.
\eea
Since $Y_U$ and $Y_D$ are diagonal, the $\mu$-term contributions
change the mass eigenvalues but do not generate any flavor mixing. The
case of the $A$-terms in more complicated. Insofar as they are
proportional to their respective Yukawa matrices, these do
not generate flavor mixing either.  But though we set $\bar{A}_{U} \propto Y_{U}$ at the high scale,
this will not be respected by the renormalization group flow as discussed previously.

At low $\tan\beta$, however, the case is moot.  With $Y_D=0$, left-right mixing in the down
sector disappears completely and in the up sector $\bar A_U \propto
\mbox{diag}(0,0,y_t)$ at all scales. Thus the LR mixing terms generate no new
flavor mixing at low $\tan\beta$.
Their sole effect is to shift the mass eigenvalues in
the top squark sector. However, since the trace of the stop mass
matrix is invariant, the sum $\tilde m^2_{t,1}+\tilde m^2_{t,2}$ is
not changed and the sum rules in Eqs.~(\ref{sum1}) and
(\ref{sum2}) remain correct. Thus sum rules which only depend on
the trace of the stop mass matrix (such as Eq.~(\ref{sum2})) rather than on its
individual eigenstates remain valid even in the presence of
non-zero left-right mixing.

So, at least for the case of low $\tan\beta$ (where $y_b$ can be ignored), we have found a way to express the flavor-changing effects in terms of the physical squark masses which may soon be available at the LHC. At low $\tan\beta$ we have seen that there are not that many masses to be measured -- just four for each rule. Of course, by ignoring the $y_b$ contributions our sum rules only work when $\tan\beta$ is low; as $\tan\beta$ increases we would expect larger and larger deviations. With that in mind, we now examine the case for large $\tan\beta$ in order to see how much more complicated (or not) it is.

\subsection{Sum Rules: The Large $\tan\beta$ Case}

At large $\tan\beta$, we can no longer ignore the effects of the
bottom Yukawa coupling on the evolution of the soft mass
parameters. In this case there are two flavor-changing matrices that
need to be
considered: $V^\dagger Y_U^2V$ as before, and now also $VY_D^2V^\dagger$. The
elements of the squark mass matrices are now:
\bea
{\cal M}^2_{U,LL} &=& (\alpha_Q+D_{u_L})\one + (\gamma_u+v_u^2) Y_U^2+
\gamma_d\,VY_D^2 V^\dagger,\nonumber\\
{\cal M}^2_{D,LL} &=& (\alpha_Q+D_{d_L})\one + \gamma_d Y_D^2+
\gamma_u V^\dagger Y_U^2V ,\nonumber\\
{\cal M}^2_{U,RR} &=& (\alpha_U+D_{u_R})\one + (2\gamma_u+v_u^2) Y_U^2, \\
{\cal M}^2_{D,RR} &=& (\alpha_D+D_{d_R})\one + 2\gamma_d Y_D^2,\nonumber
\eea
where the terms were all defined in the previous section. Notice that
we do not include the SM-like Yukawa contribution to the down squark masses which
goes as $Y_D v_d$ since this is always small, regardless of $\tan\beta$.
And as before, we
will drop the LR mixing terms for now and then reintroduce them farther along in the
calculation.

We now wish to find the mass eigenstates of the system.
In the $\tilde{u}_L$ sector, the piece of the mass matrix that generates
flavor mixing is the hermitian matrix:

\begin{eqnarray}
(\gamma_u+v_u^2)Y^{2}_{U}+\gamma_d\,VY^{2}_{D}V^{\dagger}=(\gamma_u+v_u^2)
\left(\begin{array}{ccc} \left|V_{13}\right|^2y'_{b}{}^{2} &
          V_{13}^*V_{23}y'_{b}{}^{2} & V_{13}^*V_{33}y'_{b}{}^{2} \\  &
          \left|V_{23}\right|^{2}y'_{b}{}^{2} &
          V_{23}^*V_{33}y'_{b}{}^{2} \\  &  &
          \left|V_{33}\right|^2y'_{b}{}^{2}+y_{t}^{2} \end{array}\right),
\end{eqnarray}
where we define a re-scaled bottom Yukawa:
\beq
y'_b = \left(\frac{\gamma_d}{\gamma_u+v_u^2}\right)^{1/2} y_b.
\eeq
This matrix has eigenvalues $ (\gamma_u+v_u^2)\times \{0,\frac{\epsilon
y_{t}^{2}y'{}^{2}_{b}}{y_{t}^{2}+y'_{b}{}^{2}},y_{t}^{2}
+y'_{b}{}^{2}-\frac{\epsilon
y_{t}^{2}y'{}^{2}_{b}}{y_{t}^{2}+y'_{b}{}^{2}}\} $ where $\epsilon \equiv
\left|V_{13}\right|^2+\left|V_{23}\right|^2=1-\left|V_{33}\right|^2\ll
1$.

In the down sector, the relevant matrix is $\gamma_dY_D^2+\gamma_u V^\dagger
Y_U^2 V$, which has eigenvalues identical to those above after the replacement
$(\gamma_u+v_u^2) \to \gamma_u$ and $y'_b\to y''_b$ where
\beq
y''_b = \left(\frac{\gamma_d}{\gamma_u}\right)^{1/2} y_b.
\eeq

The above diagonalization has now yielded all the squark mass
eigenvalues
\bea
(\tilde m_{u_L}^{2})_{1} & = & \alpha_Q+D_{u_L}, \nonumber \\
(\tilde m_{u_L}^{2})_{2} & = &\alpha_Q+D_{u_L} +(\gamma_u+v_u^2)
\frac{\epsilon y_t^2y'_b{}^2}{y_t^2+y'_b{}^2},
\nonumber \\
(\tilde m_{u_L}^{2})_{3} & = & \alpha_Q+D_{u_L} +(\gamma_u+v_u^2)
\left(y_t^2+y'_b{}^2-\frac{\epsilon y_t^2 y'_b{}^2}{y_t^2+y'_b{}^2}\right),
\nonumber \\
(\tilde m_{u_R}^{2})_{1,2} & = & \alpha_U+D_{u_R}, \nonumber \\
(\tilde m_{u_R}^{2})_{3} & = & \alpha_U+D_{u_R} + (2\gamma_u+v_u^2)y_t^2,
\\
(\tilde m_{d_L}^{2})_{1} & = & \alpha_Q+D_{d_L}, \nonumber \\
(\tilde m_{d_L}^{2})_{2} & = & \alpha_Q+D_{d_L} +\gamma_u
\frac{\epsilon y_t^2{y''_b}^2}{y_t^2+{y''_b}^2},
\nonumber \\
(\tilde m_{d_L}^{2})_{3} & = & \alpha_Q+D_{d_L} +\gamma_u
\left(y_t^2+{y''_b}^2-\frac{\epsilon y_t^2 {y''_b}^2}{y_t^2+{y''_b}^2}\right),
\nonumber \\
(\tilde m_{d_R}^{2})_{1,2} & = & \alpha_D+D_{d_R}, \nonumber \\
(\tilde m_{d_R}^{2})_{3} & = & \alpha_D+D_{d_R} + 2\gamma_d y_b^2. \nonumber
\eea

Now that we have solved for the mass
eigenstates we need to look at the off diagonal elements which are
related to the various mixings. As expected, the RH mixings are still
zero:
\beq
(\Delta^u_{ij})_{RR} = (\Delta^d_{ij})_{RR} = 0.
\eeq
The LL mixing in the down sector is unchanged from the low $\tan\beta$
case:
\beq
(\Delta^d_{ij})_{LL} = \gamma_u\, V^*_{3i}V_{3j} y_t^2.
\eeq
But now there are non-zero contributions to the LL mixing in the up
sector as well:
\beq
(\Delta^u_{ij})_{LL} = \gamma_d\, V^*_{i3}V_{j3} y_b^2.
\eeq

Given both mass eigenvalues and
off-diagonal elements it becomes a simple exercise in equation manipulation
to find the right combinations of masses which yield the correct
off-diagonal elements. If we take $\epsilon\to 0$ (which is almost
certainly a good enough approximation), then:
\bea
(\Delta^u_{ij})_{LL} &=&
\frac{V^*_{i3}V_{j3}}{8}
\left[3\left(\tilde m_{b_1}^{2}+\tilde m_{b_2}^{2}-\tilde m_{d_L}^{2}
-\tilde m_{d_R}^{2}\right)\nonumber \right. \\
 &   & \phantom{\frac{V^*_{i3}V_{j3}}{8}xx}
\left. -\tilde m_{t_1}^{2}-\tilde m_{t_2}^{2}+\tilde m_{u_L}^{2}
+\tilde m_{u_R}^{2}+2m_t^{2}\right], \label{highsum1} \\
(\Delta^d_{ij})_{LL} &=&
\frac{V_{3i}^*V_{3j}}{8}\left[3\left( \tilde m_{t_1}^{2}
+\tilde m_{t_2}^{2}
  -\tilde m_{u_L}^{2}-\tilde m_{u_R}^{2}
-2m_t^{2} \right)\right. \nonumber \\
 &   &\phantom{\frac{V^*_{i3}V_{j3}}{8}xx}
\left.  -\tilde m_{b_1}^{2}-\tilde m_{b_2}^{2}
+\tilde m_{d_L}^{2}+ \tilde m_{d_R}^{2}\right].\label{highsum2}
\end{eqnarray}
For the sake of completeness, we can also derive sum rules when the
$O(\epsilon)$ effects are kept:
\bea
(\Delta^u_{ij})_{LL} & = &
\frac{V^*_{i3}V_{j3}}{8}
\left[3\left(\tilde m_{b_1}^{2}+\tilde m_{b_2}^{2}+\tilde m_{s_L}^{2}
-2\tilde m_{d_L}^{2}-
  \tilde m_{d_R}^{2}\right)\nonumber \right. \\
 &   & \phantom{\frac{V^*_{i3}V_{j3}}{8}xx}
\left. -\tilde m_{t_1}^{2}-\tilde m_{t_2}^{2}-\tilde m_{c_L}^{2}
+2\tilde m_{u_L}^{2}+ \tilde m_{u_R}^{2}+ 2m_t^{2}\right],\label{highsum3} \\
(\Delta^d_{ij})_{LL} &=&
\frac{V_{3i}^*V_{3j}}{8}\left[3\left( \tilde m_{t_1}^{2}
+\tilde m_{t_2}^{2}+\tilde m_{c_L}^{2}-
  2\tilde m_{u_L}^{2}-\tilde m_{u_R}^{2}
-2m_t^{2} \right)\right. \nonumber \\
 &   &\phantom{\frac{V^*_{i3}V_{j3}}{8}xx}
\left.  -\tilde m_{b_1}^{2}-\tilde m_{b_2}^{2}-
  \tilde m_{s_L}^{2}+2\tilde m_{d_L}^{2}+
  \tilde m_{d_R}^{2}\right]\label{highsum4}.
\end{eqnarray}

Note that the only difference between Eqs.~(\ref{highsum1})--(\ref{highsum2}) and
Eqs.~(\ref{highsum3})--(\ref{highsum4}) is that we assume the difference between the first and second
generation squark masses ({\it e.g.}, $\tilde{m}^2_{s_L} -\tilde{m}^2_{d_L})$ are too small to measure in the first set.
If these differences are not small, then MFV (and mass unification) are almost certainly wrong anyways.

Finally we return to the LR mixing sector, which
affects us in two ways: first, shifting the mass eigenvalues in a way
which has no automatic correlation with the LL mixing, and second,
generating explicit LR flavor-mixing insertions. As in the low $\tan\beta$ case, the first effect can
introduce a correction to, or even disrupt,
the previous LL sum rules; the second is all
important in deriving sum rules for $(\Delta^{u,d}_{ij})_{LR}$.

First, what is the effect of LR mixing on the LL sum rules?
In building the LL sum rules,
we were careful to include only the stop and sbottom
masses in the combinations which represent the trace over the squark mass
matrices. For low $\tan\beta$, the only non-zero LR term (in the superCKM
basis) was the $\tilde t_L-\tilde t_R$ mass term, and so writing our
sum rules in terms of $m^2_{\tilde t_L}+m^2_{\tilde t_R}$
was enough to guarantee that the
sum rules survived even after LR mixing was reintroduced.

At large $\tan\beta$, this is not necessarily enough. In both the up
and down sectors, the $A$-terms receive RGE corrections that are
neither diagonal nor small:
\begin{eqnarray}
\delta \bar A_U &\propto& 3A_0VY_D^2V^\dagger Y_U \nonumber \\
\delta A_D &\propto & 3A_0V^\dagger Y_U^2 V Y_D
\end{eqnarray}
in the superCKM basis. The largest contributions from these terms are
still to the (3,3) element, but there are non-neglible contributions
to the ($i$,3) elements as well:
\begin{eqnarray}
\delta (\bar A_U)_{i3} &\propto& 3A_0 y_t y_b^2 V_{tb}^* V_{ib}\nonumber \\
\delta (A_D)_{i3} &\propto & 3A_0 y_t^2 y_b V_{tb} V_{ti}^*.
\end{eqnarray}
By how much do these contributions alter the previous sum rules?
Luckily, by very little. The shift in the eigenvalues of the $u$-squark
mass matrix is at most $O(V_{cb})$, and in the $d$-squark
matrix it is at most $O(V_{ts})$.
That is, the shifts in the masses due to the
flavor-changing LR terms are highly suppressed compared to the
dominant LL flavor-diagonal terms. We have checked numerically that dropping these small terms makes almost no difference in the validity of the sum rules.

What about the explicit $(\Delta^{u,d}_{LR})_{ij}$ elements? First,
in the $d$-squark sector one must recall that the $A$-term
contributions are suppressed by $1/\tan\beta$ compared to the LL and
RR mass terms, and so can be ignored. The $\mu$-term contributions are
not suppressed but are flavor-diagonal in the superCKM basis. The only
sizable element is flavor conserving, $(\Delta^d_{LR})_{33}$, and is
discussed further below.

In the $u$-squark sector, it is the $\mu$-term contributions that are
suppressed by $1/\tan\beta$ and so can be ignored. But that leaves the
$\bar A_U$ terms present as discussed above. Because the RGEs only
generate terms of the form $(\Delta^u_{LR})_{i3}$ we have:
\beq
(\Delta^{u,d}_{i1})_{LR}=(\Delta^{u,d}_{i2})_{LR}=0\quad (i=1,2,3).
\eeq

We are also in need of sum rules for $(\Delta^u_{i3})_{LR}$ for
$i=1,2$. However such sum rules would be both very difficult to use
and rather useless. They are difficult to use because they would
necessarily involve measuring left-right mixing in the first and
second generation $u$-squarks. They are anyway useless because there
are two competing contributions to LR amplitudes, the explicit
$(\Delta^u_{i3})_{LR}$ insertion and the double-insertion
$(\Delta^u_{i3})_{LL}(\Delta^u_{33})_{LR}/\tilde m^2$.
For most cases, the double
insertion is comparable to (and usually larger than) the single
insertion. Because their relative sizes and phases are model-dependent,
it would be impossible to describe the total LR contribution
in the nice, closed form of sum rule.

While they can not produce FCNCs a sumrule for $(\Delta^u_{LR})_{33}$
already exists and can be found by Martin and Ramond ~\cite{Martin:1993ft} for
$y_b=0$, but
which is here given for non-zero $y_b$:
\begin{eqnarray}
(\Delta^u_{33})_{LR} & = &
  \frac{1}{2}\left[\left(m_{\tilde{t}_{1}}^{2}-m_{\tilde{t}_{2}}^{2}
  \right)^{2} \nonumber \right. \\  &   & \left.
  -\left(m_{\tilde{u}_{L}}^{2}-m_{\tilde{u}_{R}}^{2}+
  \frac{1}{2}\left(m_{\tilde{b}_{1}}^{2}+m_{\tilde{b}_{2}}^{2}
  -m_{\tilde{t}_{1}}^{2}-m_{\tilde{t}_{2}}^{2}\right)+m_{\mbox{t}}^{2}
  \nonumber \right. \right.  \\  &   & \left. \left.
  -\frac{1}{2}\left(m_{\tilde{d}_{L}}^{2}+m_{\tilde{d}_{R}}^{2}
  -m_{\tilde{u}_{L}}^{2}-m_{\tilde{u}_{R}}^{2}\right)\right)^{2}
  \right]^{\frac{1}{2}}.
\end{eqnarray}

Similarly a  sum rule for $(\Delta^d_{33})_{LR}$ can be derived it is
giving below.
\begin{eqnarray}
(\Delta^d_{33})_{LR} & = &
  \frac{1}{2}\left[\left(m_{\tilde{b}_{1}}^{2}-m_{\tilde{b}_{2}}^{2}
  \right)^{2} \nonumber \right. \\  &   & \left.
  -\left(m_{\tilde{d}_{L}}^{2}-m_{\tilde{d}_{R}}^{2}+
  \frac{1}{2}\left(m_{\tilde{t}_{1}}^{2}+m_{\tilde{t}_{2}}^{2}
  -m_{\tilde{b}_{1}}^{2}-m_{\tilde{b}_{2}}^{2}\right)-m_{\mbox{t}}^{2}
  \nonumber \right. \right.  \\  &   & \left. \left.
  -\frac{1}{2}\left(m_{\tilde{u}_{L}}^{2}+m_{\tilde{u}_{R}}^{2}
  -m_{\tilde{d}_{L}}^{2}-m_{\tilde{d}_{R}}^{2}\right)\right)^{2}
  \right]^{\frac{1}{2}}.
\end{eqnarray}

This term really only matters in the large $\tan\beta$ limit where the
$\mu$ dominates.  Once $\mu$ and $\tan\beta$ are determined from other
sources this sumrule will provide another test of the model.

\section{Application: $b\to s\gamma$}

Throughout this paper we are considering models which exhibit minimal flavor violation because
their squark masses unify at some scale and because the squark mass splittings and mixings are
induced by the Yukawa couplings alone. This particular class of SUSY models is attractive for several
reasons, but chief among them is that they reduce the level of flavor changing in the MSSM to be consistent
with experimental bounds. Within MFV models, many observables become rather insensitive to gluino-induced
flavor changing; meson-anti-meson mixing is a particularly good example here. Yet there remain FCNC processes
in which gluino-mediated amplitudes can still generate measurably large deviations. Chief among
these is the decay $b\to s\gamma$.

The rare decay $b\to s\gamma$ is generated by two dimension-5 operators: $\bar b_R \sigma_{\mu\nu} s_L F^{\mu\nu}$
and $\bar b_L \sigma_{\mu\nu} s_R F^{\mu\nu}$. Both processes require a chirality flip which introduces,
by virtue of the chiral symmetries of the SM, at least one factor of $m_b$ and $m_s$ respectively in the
coefficients for each. Thus the operators are effectively dimension-6 in terms of the heavy
(weak) mass scale, suppressed by $m_{b,s}/M_W^2$. Because $m_b\gg m_s$ we will work only with the first
operator for the remainder of this section.

The SM contribution to $b\to s\gamma$ is even further suppressed by a
GIM cancellation among the various up-type quarks which run in the loop. In all, it yields an amplitude
 ~\cite{Gabbiani:1988rb} :
\bea
A(b\to s\gamma)& \propto & \frac{m_b}{M^2_W} \sum^3_{k=1} V_{3k} V_{k2}^* F(m_{u_k}),
\eea
where $F(m)$ is a kinematic function of the up-quark masses.
Clearly this amplitude is suppressed due to the smallness of the CKM off-diagonal elements
and would be identically zero if the quark masses were all equal.

But as a 1-loop, helicity-suppressed process within the SM, new physics contributions can generate sizable corrections.
SUSY in particular is well known for its many and varied contributions to $b\to s\gamma$. These contributions
come from penguin diagrams mediated by internal charginos ($\tilde{\chi}^{\pm}$), charged Higgs bosons ($\mbox{H}^{\pm}$)
or gluinos ($\tilde{g}$). Even though the mass scale associated with each of these is probably higher than $M_W$,
 these contributions can actually dominate over the SM. For one thing, with an internal Higgs or
higgsino, the external $m_b$ mass flip can be replaced with an internal $y_b$ Yukawa factor, which is enhanced over
$m_b$ by roughly $1/\cos\beta$. Second, for internal gluons and neutralinos, their
Majorana nature allows a spin-flip inside the diagram,
again removing the $m_b$ suppression that the SM diagram exhibits.

In most discussions of the supersymmetric contributions to $b\to s\gamma$, only the chargino and charged Higgs
contributions are considered. Both of these contributions can be individually large, though there is typically some
cancellation among the various pieces. (In unbroken SUSY, there are no magnetic moment transitions, and therefore
no $b\to s\gamma$.)

However, in generic, non-MFV models, it is the gluino diagrams that are expected to dominate this process thanks to the
large $\alpha_s$. Because large deviations from the SM have not been observed, strict limits on $(\Delta_{32})_{LL}$
are derived for such models. In MFV models, on the other hand, the gluino diagrams are suppressed by the
(super)GIM mechanism to a level that they are usually ignored in discussions of $b\to s\gamma$.
But we will show in the discussion below that in some parts of the SUSY parameter space, the gluino diagrams can be
the same size, or larger, than the usual chargino and charged Higgs contributions, and that the sum rules are
particularly well-suited for disentangling the gluino contribution.

The diagrams which dominate the gluino contribution are shown in
Fig.~1.
The first diagram includes a single
$(\Delta^d_{32})_{LL}$ insertion to change flavor on the squark line, but is suppressed by $m_b$ due to a
helicity flip on the external $b$-quark line.  The second diagram requires a double insertion $(\Delta^d_{33})_{LR} (\Delta^d_{32})_{LL}$
on the squark line but gets its helicity flip by an internal gluino mass insertion. This,
coupled with the fact that $(\Delta^d_{33})_{LR} \approx \mu \tan\beta$, allows the gluino contributions
to become comparable to the chargino and charged Higgs contributions at large $|\mu|$ and large $\tan{\beta}$.

\begin{figure}[t]
\centering
\epsfxsize=4in
\hspace*{0in}
\epsffile{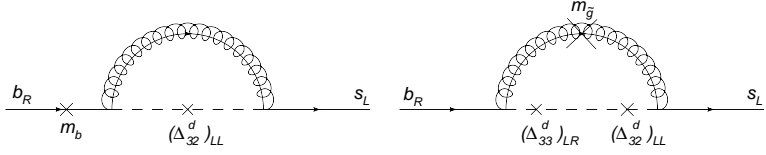}
\caption{Gluino contributions to $b\to s\gamma$.}
\end{figure}

In order to parametrize the effects of SUSY on $b\to s\gamma$, we will use the standard Wilson operator expansion:
\bea
{\cal H}(B\to X_s\gamma) = \sum_{i=1}^8 C_i {\cal O}_i
\eea
where $C_i$ are the Wilson coefficients of the operators ${\cal O}_i$.  For this paper we will focus on the short
distance effects contained in the Wilson coefficients, with our attention devoted to the two operators:
\begin{eqnarray*}
{\cal O}_7&=&m_b \bar{s}_L\sigma^{\mu\nu} b_R F_{\mu\nu} \\
{\cal O}_8&=&m_b \bar{s}_L\sigma^{\mu\nu} T^a b_R G_{\mu\nu}^a
\end{eqnarray*}
We will define variables $r_{7,8}$ to parametrize the size of the SUSY contributions relative to the SM:
\bea
r_7& = & \frac{C_{7,SM}+C_{7,SUSY}}{C_{7,SM}},\nonumber\\
r_8& = & \frac{C_{8,SM}+C_{8,SUSY}}{C_{8,SM}}. \label{r}
\eea
where, for simplicity, we will take the $C_i$'s to be real (we will not consider any CP violation in this discussion).
The contributions to the Wilson coefficient $C_7$ from $W$-bosons,
charged Higgs, charginos and gluinos are given below~\cite{Bertolini:1990if,Barbieri:1993av}:
\begin{eqnarray}
C_{7,SM} & = & -\frac{3 \alpha_w \sqrt{\alpha_s}}{8 \sqrt{\pi}} \frac{1}{M^2_W} V^*_{ts} V^*_{tb}x_{tW}
\left[\frac{2}{3} F_1(x_{tW})+F_2(x_{tW})\right],\nonumber\\
C_{7,H^\pm}&=&-\frac{\alpha_w \sqrt{\alpha}}{8\sqrt{\pi}}\frac{1}{M^2_W}V^*_{ts}V_{tb}x_{tH^{\pm}}\times\nonumber\\
&&\left(\frac{1}{\tan^2\beta}\left[\frac23 F_1(x_{tH^{\pm}})+F_2(x_{tH^{\pm}})\right]+
\frac23 F_3(x_{tH^{\pm}})+F_4(x_{tH^{\pm}})\right),\nonumber\\
C_{7,\tilde{\chi}^{\pm}}&=&\frac{\alpha_w\sqrt{\alpha}}{4\sqrt{\pi}}\frac{1}{M^2_W}V^*_{ts}V_{tb}
\sum^2_{j=1}\left[x_{W\tilde{\chi}_j}\left[\left|{\cal{V}}_{j1}\right|^2\left(-\frac23 F_1(x_{\tilde{q}\tilde{\chi}_j})-
F_2(x_{\tilde{q}\tilde{\chi}_j})\right)\right.\right.\nonumber\\&&\left.\left.-\sum^2_{k=1}\left|
{\cal{V}}_{j1}{\cal{T}}_{k1}-{\cal{V}}_{j2}{\cal{T}}_{k2}\frac{m_t}{\sqrt{2}M_W\sin\beta}\right|^2\left
(-\frac23 F_1(x_{\tilde{t}_k\tilde{\chi}_j})-
F_2(x_{\tilde{t}_k\tilde{\chi}_j})\right)\right]\right.\nonumber\\&&\left.-\frac{{\cal{U}}_{j2}M_w}{\sqrt{2}
m_{\tilde{\chi}^{\pm}}\cos\beta}\left[{\cal{V}}_{j1}F_5(x_{\tilde{q}\tilde{\chi}_j})\right.\right.\nonumber\\&&\left.\left.
-\sum^2_{k=1}\left({\cal{V}}_{j1}{\cal{T}}_{k1}-{\cal{V}}_{j2}{\cal{T}}_{k2}\frac{m_t}{\sqrt{2}M_W\sin\beta}
\right){\cal{T}}_{k1}F_5(x_{\tilde{t}_k\tilde{\chi}_j})\right]\right],\\
C_{7,\tilde{g}} & = &  -\frac{\alpha_s \sqrt{\alpha}}{\sqrt{\pi}}\frac{2}{9}\sum_{k=1}^{6}\frac{1}{m^2_{d_k}}
\left( \Gamma^{kb}_{DL} \Gamma^{*ks}_{DL} F_2(x_{\tilde{g}\tilde{d}_k})-
\Gamma^{kb}_{DR} \Gamma^{*ks}_{DL} \frac{m_{\tilde{g}}}{m_b} F_4(x_{\tilde{g}\tilde{d}_k})\right).\nonumber
\end{eqnarray}
The coefficient $C_8$ receives a similar set of contributions:
\begin{eqnarray}
C_{8,SM} & = & -\frac{\alpha_w \sqrt{\alpha_s}}{4\sqrt{\pi}} \frac{1}{M^2_W} V^*_{ts} V_{tb}x_{tw}
F_1(x_{tW}),\nonumber\\
C_{8,H^\pm}&=&-\frac{\alpha_w \sqrt{\alpha_s}}{8\sqrt{\pi}}\frac{1}{M^2_W}V^*_{ts}V_{tb}x_{tH^{\pm}}\times\nonumber\\
&&\left(\frac{1}{\tan^2\beta}F_1(x_{tH^{\pm}})+F_3(x_{tH^{\pm}})\right),\nonumber\\
C_{8,\tilde{\chi}^{\pm}}&=&\frac{\alpha_w\sqrt{\alpha_s}}{4\sqrt{\pi}}\frac{1}{M^2_W}V^*_{ts}V_{tb}
\sum^2_{j=1}\left[x_{W\tilde{\chi}_j}\left[\left|{\cal{V}}_{j1}\right|^2\left(-F_1(x_{\tilde{q}\tilde{\chi}_j})
\right)\right.\right.\nonumber\\&&\left.\left.\sum^2_{k=1}\left|
{\cal{V}}_{j1}{\cal{T}}_{k1}-{\cal{V}}_{j2}{\cal{T}}_{k2}\frac{m_t}{\sqrt{2}M_W\sin\beta}\right|^2F_1(x_{\tilde{t}_k\tilde{\chi}_j})
\right]\right.\nonumber\\&&\left.-\frac{{\cal{U}}_{j2}M_w}{\sqrt{2}
m_{\tilde{\chi}^{\pm}}\cos\beta}\left[{\cal{V}}_{j1}F_6(x_{\tilde{q}\tilde{\chi}_j})\right.\right.\nonumber\\&&\left.\left.
-\sum^2_{k=1}\left({\cal{V}}_{j1}{\cal{T}}_{k1}-{\cal{V}}_{j2}{\cal{T}}_{k2}\frac{m_t}
{\sqrt{2}M_W\sin\beta}\right){\cal{T}}_{k1}
F_6(x_{\tilde{t}_k\tilde{\chi}_j})\right]\right],\\
C_{8,\tilde{g}} & = &  -\frac{\alpha_s \sqrt{\alpha_s}}{4\sqrt{\pi}} \sum_{k=1}^{6}\frac{1}{m^2_{d_k}}
\left(\Gamma^{kb}_{DL}\Gamma^{*ks}_{DL}\left(3F_1(x_{\tilde{g}\tilde{d}_k})+\frac{1}{3}
F_2(x_{\tilde{g}\tilde{d}_k})\right)\right.\nonumber\\
&&\left.-\Gamma^{kb}_{DR}\Gamma^{*ks}_{DL}\frac{m_{\tilde{g}}}{m_b}
\left(3F_3(x_{\tilde{g}\tilde{d}_k})+\frac{1}{3}F_4(x_{\tilde{g}\tilde{d}_k})\right)\right),\nonumber
\end{eqnarray}
where ${\cal{U}}$ and ${\cal{V}}$ are the chargino mixing matrices and ${\cal{T}}$
is the the top squark mixing matrix. The functions $F$ are loop functions and are
defined using $x_{ij} = m^2_{i}/ m^2_{j}$ to be:
\bea
F_1(x)& = & \frac{1}{12(x-1)^4}(x^3-6x^2+3x+2+6x\log{x}),\nonumber\\
F_2(x)& = & \frac{1}{12(x-1)^4}(2x^3+3x^2-6x+1-6x^2\log{x}),\nonumber\\
F_3(x)& = & \frac{1}{2(x-1)^3}(x^2-4x+3+3\log{x}),\nonumber\\
F_4(x)& = & \frac{1}{2(x-1)^3}(x^2-1+2x\log{x}),\nonumber\\
F_5(x)& = & (x-1)\left[\frac23 F_1(x)+F_2(x)\right]+\frac{x}{2}\left[\frac23 F_3(x)+F_4(x)\right]-\frac{23}{36},\nonumber\\
F_6(x)& = & (x-1)F_1(x)+\frac{x}{2}F_3(x)-\frac{1}{3}.\nonumber
\eea
Finally the $\Gamma$ matrices are the matrices which diagonalize the $6\times6$ down squark mass matrix.  These are
defined such that:
\bea
M_{\tilde{d}}^2(diag)& = &\Gamma_{D}^{\dagger} M_{\tilde{d}}^2 \Gamma_{D},
\eea
with the $\Gamma^{kj(j+3)}_{DL(R)}$ designating a transition from the $k$th down type squark in a mass eigenstate to
the $j$th left-handed (right-handed) down squark in the interaction basis.

For the non-gluino contributions shown above, we have completed our numerical calculations using a modification of the $b\to s\gamma$ routines from the program SPheno~\cite{Porod:2003um}. To these routines we have added corrections coming from finite SUSY loops that shift the $b$-quark Yukawa coupling away from its tree-level value, which are only important at large $\tan\beta$~\cite{Carena:2000uj}.

For the gluino diagrams, the $\Gamma$ matrices in the above formulae contain the information concerning  $(\Delta^d_{32})_{LL}$
and $(\Delta^d_{33})_{LR}$; while easy enough to use numerically, they are
difficult to use analytically. Since the two biggest contributions are going to be
$(\Delta^d_{32})_{LL}$ and $(\Delta^d_{33})_{LR}$ we can simplify the expressions for for $C_{i,\tilde{g}}$
considerably.

\begin{eqnarray}
C_{7,\tilde{g}}&=& \frac{\alpha_s \sqrt{\alpha}}{\sqrt{\pi}}\frac{4}{9}
\left[\frac{(\Delta^d_{32})_{LL}}{m^2_{\tilde{s}_L}-m^2_{\tilde{b}_L}}\left(\frac{1}{m^2_{\tilde{s}_L}}
F_2(x_{\tilde{g}\tilde{s}_L})-
\frac{1}{m^2_{\tilde{b}_L}}F_2(x_{\tilde{g}\tilde{b}_L})\right)\right.\nonumber\\
& & \left. +\frac{1}{m^2_{\tilde{b}_L}}\frac{m_{\tilde{g}}}{m_b}
\frac{(\Delta^d_{32})_{LL}}{m^2_{\tilde{s}_L}-m^2_{\tilde{b}_L}}
\frac{(\Delta^d_{33})_{LR}}{m^2_{\tilde{b}_L}-m^2_{\tilde{b}_R}}F_4(x_{\tilde{g}\tilde{b}_L})\right],\nonumber\\
C_{8,\tilde{g}} & = & \frac{\alpha_s \sqrt{\alpha_s}}{2\sqrt{\pi}}\left[\frac{(\Delta^d_{32})_{LL}}
{m^2_{\tilde{s}_L}-m^2_{\tilde{b}_L}}\left(\frac{1}{m^2_{\tilde{s}_L}}(3 F_1(x_{\tilde{g}\tilde{s}_L})
+\frac{1}{3}F_2(x_{\tilde{g}\tilde{s}_L}))\right.\right.\nonumber\\
&&\left.\left.-
\frac{1}{m^2_{\tilde{b}_L}}\left(3 F_1(x_{\tilde{g}\tilde{b}_L})+\frac{1}{3} F_2(x_{\tilde{g}\tilde{b}_L})\right)
\right)\right. \nonumber\\
& & \left. -\frac{1}{m^2_{\tilde{b}_L}}\frac{m_{\tilde{g}}}{m_b}
\frac{(\Delta^d_{32})_{LL}}{m^2_{\tilde{s}_L}-m^2_{\tilde{b}_L}}
\frac{(\Delta^d_{33})_{LR}}{m^2_{\tilde{b}_L}-m^2_{\tilde{b}_R}}
\left(3F_3(x_{\tilde{g}\tilde{b}_L})+\frac{1}{3}F_4(x_{\tilde{g}\tilde{b}_L})\right)\right],
\end{eqnarray}
where only contributions from $\tilde{s}_L, \tilde{b}_L \mbox{ and } \tilde{b}_R$ are
included in the loops. We leave the above equations in terms of $\tilde{b}_R$ and $\tilde{b}_L$ instead of
$\tilde{b}_1$ and $\tilde{b}_2$ for the sake of simplicity. Since the equations above require the dominant entries in the
$\Gamma_D$, going from $\tilde{b}_R$ and $\tilde{b}_L$ to $\tilde{b}_1$ and $\tilde{b}_2$ is trivial.

With the Wilson coefficients in hand, it is time to focus on the long distance effects.
These effects in $b\to s\gamma$ have been studied in great detail over the years, see Refs.~\cite{Hurth:2003dk,Misiak:2006zs} for the most recent and complete discussion.
We use the results from Ref.~\cite{Hurth:2003dk} which has NLL result presented
with clean analytical formulae that easily incorporate new physics by using the ratios
$r_7 \mbox{ and } r_8$ defined in Eqs.~(\ref{r}).  In discussions of the
long distance effects, a large systematic uncertainty always comes in from the ratio $m_c/m_b$.
Following Ref.~\cite{Domingo:2007dx}, we use this uncertainty to tune
the ratio to $0.31$, which produces a SM prediction of Br($\bar B\to X_s\gamma$) more closely in line with the current
NNLO theoretical value of $(3.15 \pm 0.23)\times10^{-4}$ ~\cite{Misiak:2006zs}.

The ratios $r_7 \mbox{ and } r_8$ then can be used to acquire
$\mbox{Br}(\bar{B}\to X_s \gamma)$ through the following equation,
\begin{equation}
\mbox{Br}(\bar{B} \to X_s \gamma)  =  \frac{N}{100} \left|{\frac{V^*_{ts}V_{tb}}{V_{cb}}}\right|^2B^{unn}
\end{equation}
where $B^{unn}$ is the ``un-normalized" branching fraction given by:
\begin{eqnarray}
B^{unn}  &=&   6.7603 + .8161 r_7^2 + 4.517 r_7 + 0.0197 r_8^2 + .5427 r_8\nonumber\\
        &   &  + .3688\left|{\epsilon_s}\right|^2 +  -2.82953 \mbox{Re}(\epsilon_s)+2.96158 \mbox{Im}(\epsilon_s)\nonumber\\
        &   &  + .1923 (r_8 r_7) + -1.146 r_7 \mbox{Re}(\epsilon_s) + -.0855 r_8 \mbox{Re}(\epsilon_s)\nonumber\\
        &   &  + -1.0677 r_7 (-\mbox{Im}(\epsilon_s)) + -0.0799 r_8 (-\mbox{Im}(\epsilon_s)).
\end{eqnarray}
In the equation above, $N=2.567(1\pm0.064)\times10^{-3}$ is the normalization factor and
\bea
\epsilon_s & = & \frac{V^*_{us} V_{ub}}{V^*_{ts} V_{tb}}=(-0.088\pm 0.0024)+ i(0.0180\pm 0.0015).\nonumber
\eea

\begin{figure}[t]
\centering
\epsfxsize=4in
\hspace*{0in}
\epsffile{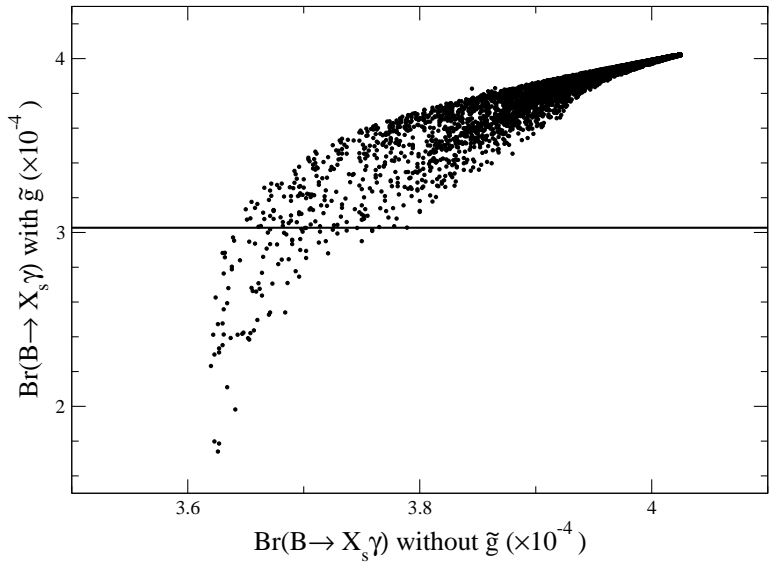}
\caption{Plot of Br($\bar{B}\to X_s \gamma$) with and without the $\tilde{g}$ contributions.
The horizontal line on the graph represents the $2\sigma$ experimental lower limit. These points all have $m_{H^\pm}=300\gev$, $\mu=1\tev$ and $\tan\beta=30$.}
\label{fig2}
\end{figure}

In order to compare the strength of gluino diagrams to the usual charged Higgs and chargino contributions, it is helpful to have a set of models in hand. For that purpose,
we scan the parameter space of the MSSM, calculating
Br$(\bar{B} \to X_s \gamma)$ both with and without the gluino contribution.
For ease of calculation we have chosen the canonical mSUGRA boundary conditions for the squark and
gaugino sectors, applied at the gauge coupling unification scale.  This will guarantee that the parameter space we are exploring
is minimally flavor violating. Since our interest lies in MFV models we have a little more freedom with the Higgs sector
which was not forced to unify with the other masses. This allows us to treat $\mu$ and $m_{H^{\pm}}$ as free parameters.  A random sampling of this parameter space is shown in Figure~\ref{fig2}, where we have allowed the mSUGRA parameters to vary over the ranges:
\begin{eqnarray}
100\gev <&  m_0 & <1000 \gev\nonumber\\
100\gev <&  m_{1/2}  & <1000 \gev\nonumber\\
-500\gev <& A_0 &< \,\,500 \gev,\nonumber
\end{eqnarray}
while we set $m_{H^\pm}=300\gev$, $\mu=1\tev$ and $\tan\beta=30$.
In the figure we present a comparison of the calculated Br($\bar B\to X_s\gamma$) rate with and without the gluino contributions among the other SUSY diagrams. Specifically, we show along the $x$-axis a calculation of B($\bar B\to X_s\gamma$) including only the chargino, charged Higgs and SM contributions for model points in the ranges defined above; along the $y$-axis we show the exact same models, but now including the gluinos in the $b\to s\gamma$ amplitude.

In order to create this plot we calculated the squark and gaugino spectra using the 1-loop renormalization group equations. The Higgs masses were determined using the highly precise calculation encoded into CPSuperH~\cite{Lee:2003nta}. So that non-physical models, or models already ruled out experimentally, were not included among the points in the figure, we applied a set of cuts to the parameter space; specifically, we required that the lightest neutralino be the lightest SUSY particle (LSP), $m_{\tilde t_1} > 96\gev$, $m_{\tilde \chi^\pm_1} > 103\gev$ and $m_h > 90 GeV$.

The light Higgs mass bound is the most complicated. The bound for a SM-like Higgs is $114\gev$~\cite{Barate:2003sz}, but can be lower in extensions of the SM, especially those with additional light Higgs fields. We have examined our results from Figure~\ref{fig2} for tighter cuts on the Higgs mass, all the way up to $114\gev$. We find that the points with the largest gluino contributions to $b\to s\gamma$ tend to have the lightest Higgs masses, and are therefore cut out of the parameter space as the Higgs mass constraint is tightened. The reason is simple: light Higgs masses well above the $Z$-mass require large top squark masses. But because of the mSUGRA boundary conditions, this drives the bottom squark masses to also be heavy, and
these in turn force the gluino loops in $b\to s\gamma$ to decouple. One way to satisfy both requirements is, for example, to modify the mSUGRA boundary conditions so that the $\tilde t_R$ becomes heavy separately from the other squarks, which pushes up the Higgs mass without directly affecting the gluino loop in $b\to s\gamma$. We examined models in which $m_{\tilde U,0}$ was varied independently of the other squark masses, and the effect of the Higgs mass constraint was essentially eliminated.

One last constraint applied to the points in the figure is that one or the other calculation of Br$(\bar B\to X_s\gamma)$ has so fall within the experimental 95\% confidence region\footnote{We use the Heavy Flavor Averaging Group's world average of Br$(\bar B\to X_s\gamma) = (3.55\pm 0.24^{+0.09}_{-0.10}\pm 0.03)\times 10^{-4}$ for $E_\gamma>1.6\gev$~\cite{Barberio:2006bi}.}: $(3.03\div
4.07)\times  10^{-4}$. Points in which both calculations (with and without gluinos) fall outside that range are eliminated, but points in which one or the other calculation falls within the range are kept. This allows us to see quite clearly that the effect of the gluino contributions can be quite large, though its sign is always the same: the branching fraction with the gluinos included is always lower than that without the gluino. (We will discuss the reasons for this in detail below.) Thus the gluino contributions tend to rule out models which would otherwise appear to be consistent with experiment, tightening constraints in the parameter space.

\begin{figure}[t]
\centering
\epsfxsize=4in
\hspace*{0in}
\epsffile{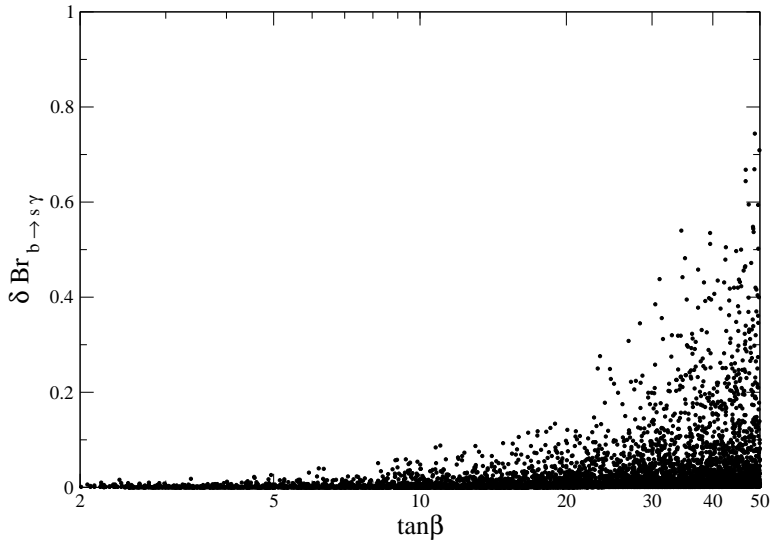}
\caption{Plot of $\delta$Br$(\bar B\to X_s\gamma)$ vs.\ $\tan\beta$ normalized by the SM theoretical ratio of $3.15\times 10^{-4}$}
\label{fig3}
\end{figure}

In Figure~\ref{fig3} we have shown the same set of points in a different way, in order to emphasize the magnitude and sign of the gluino effect. Here we plot $\tan\beta$ versus the normalized difference in the two calculations of $\bar B\to X_s\gamma$. Specifically, we plot along the $y$-axis the quantity:
\begin{eqnarray}
\delta\, \mbox{Br}_{\,b\to s\gamma} & = &\frac{\mbox{Br}(\bar B\to X_s\gamma)_{\rm no\,\, \tilde{g}} - \mbox{Br}(\bar B\to X_s\gamma)_{\rm with\,\,\tilde{g}}}
{\mbox{Br}(\bar B\to X_s\gamma)_{\rm SM}},
\end{eqnarray}
normalized by the SM branching ratio given to be $3.15\times10^{-4}$. The figure again shows the the decrease of the branching ratio due to the gluino contributions, but also the strong dependence on $\tan\beta$, which is expected since we require large $\tan\beta$ in order to generate significant LR mixing. We could also have plotted $|\mu|$ on the $x$-axis, but the shape would have been identical, since the LR mixing insertion is of the form $\mu\tan\beta$.

Notice in the figure, again, that the gluino contributions always pull down the rate for $b\to s\gamma$ (and by as much as 50\% in many cases). When $\mu$ is positive, the gluino diagrams have the opposite sign from the $W^\pm$ and $H^\pm$ diagrams, having the effect of partially canceling out the $H^\pm$ piece and pulling the branching ratio more in line with experiment.  But since the sign of the gluino contribution is pegged to the sign of $\mu$, one should expect to see the opposite behavior for $\mu<0$. In fact, this is the case, though it cannot be seen from the figure. For $\mu<0$ and $\tan\beta$ large we find all points, with and without the gluinos included, to be above the experimental limit. Thus only the $\mu>0$ case is important for us, and so we always see it destructively interfere with the SM and charged Higgs contributions.

The first conclusion we should draw from these results is that is $\tan\beta$ happens to be large, then it is quite possible for gluino diagrams to contribute significantly to the rate for $b\to s\gamma$. This is true even in the models where one would least expect it, namely MFV models. This contribution is almost uniformly ignored in the literature and has the effect of ruling out models which might otherwise have been thought to be consistent with current experimental bounds.

Second, if the gluino contribution is non-negligible, then it should be possible to extract it from the data, which would in turn provide us with an avenue for measuring $(\Delta^d_{32})_{LL}$. Of course extracting the gluino contribution from the data requires a careful measurement of the physical masses and mixings that enter the MSSM calculation of $b\to s\gamma$. This will represent quite a challenge, since it requires the measurement of chargino and top squark mixing angles. The gluino contribution also requires as input the bottom squark mixing (in the guise of $(\Delta^d_{33})_{LR}$) if we hope to extract the 3-2 mixing angle. Nonetheless, once sufficient measurements of gaugino and squark masses and mixings have been made, the problem of extracting the inter-generational mixing will become important, because it will be one of the few tests we will have for the underlying flavor independence of the SUSY-breaking sector. In particular, having a set of tests (or even a single test) of the MFV scenario will be of great importance. Since the sum rules explicitly test minimal flavor violation, and do so in a way that is independent of the exact scale or nature of the mass unification, makes them a valuable tool for doing just that.

Finally, what of other tests of MFV other than $b\to s\gamma$? We have examined a number of other FCNC's for sensitivity to minimal flavor violation and for usefulness of the sum rule approach. Though the sum rules can be quite good at extracting the mixings required to calculate gluino contributions to $K^0$-$\bar K^0$ and $B_{(s)}^0$-$\bar B_{(s)}^0$ mixing, we found the resulting contributions to be far too small to be interesting, at least in motivated models. There is also a well-known and large contribution of neutral Higgs bosons to $B_{(s)}^0$-$\bar B_{(s)}^0$ mixing~\cite{Buras:2002wq}, but at this time we can find no easy way to correlate those sources of flavor change to a set of sum rules. The best hope for extracting the new contributions to this process are by comparison to $B_s\to\mu\mu$ and $B\to X_s\mu\mu$ rates.

\section*{Conclusion}

Though SUSY solves or alleviates a number of important problems within the structure of the Standard Model, it does so at a price. That price is the flavor problem, and it is this problem that has driven most of the model-building within the SUSY community for the last twenty years. With few exceptions (such as decoupling models), solutions to the flavor problem have generally fallen into the broad class of minimal flavor violation in which all quark flavor violation is tied to the Yukawa couplings.

In minimally flavor-violating models, there are still FCNCs, including those mediated by neutral particles such as neutralinos and gluinos. They are just suppressed by the unitarity of the CKM matrix and the near-degeneracy of the squarks. But the degeneracy is broken by the Yukawa couplings themselves, which re-introduces the FCNCs. Unfortunately the effects are small and difficult to extract directly from measurements at either the LHC or even some future linear collider. It is up to high-precision flavor experiments to measure the rates and processes that will allow extraction of the details of the SUSY flavor sector.

In this paper we derived a set of sum rules that can be used to extract the flavor mixing from the masses of the squark mass eigenstates, assuming minimal flavor violation. These sum rules provide a consistency check on minimal flavor violation. Even if the low-energy spectrum appears to be consistent with some kind of mass unification, the sum rules can be used to check this explicitly.

Finally, we showed that the classic FCNC decay $b\to s\gamma$ may be a particularly good place to look for these MFV contributions, via the gluino-mediated diagrams. Though the gluino contributions are often overlooked, at large $\tan\beta$ they may actually contribute enough to change the branching fraction by 50\%. In such a case, it will be necessary to calculate the size of the $\tilde s_L$--$\tilde b_L$ flavor mixing, which is a job well-suited to the sum rules.

Once the mass spectrum of the MSSM is measured, assuming it is, it is questions about the flavor mixing that will ultimately help us disentangle the nature of the SUSY-breaking mechanism. Tools such as the sum rules, which connect flavor mixing to squark masses, could be key elements in this process.

\section*{Acknowledgments}
This work was partially supported by the National Science Foundation under grant PHY-0355066 and by the Notre Dame Center for Applied Mathematics.

\end{document}